\documentclass{LMCS}

\def\dOi{11(1:15)2015}
\lmcsheading%
{\dOi}
{1--23}
{}
{}
{Jan.~\phantom06, 2014}
{Mar.~31, 2015}
{}

\ACMCCS{[{\bf Theory of computation}]: Logic---Verification by model checking}


\usepackage{amsmath,amsfonts,amssymb,amsthm}
\usepackage[long]{macros}
\usepackage{url}
\usepackage{microtype}
\usepackage{colonequals}
\usepackage{xspace}
\usepackage{tikz}
\usepackage{booktabs}
\usepackage{numprint}
\usepackage{hyperref}

\npthousandsep{\,}
\npthousandthpartsep{}
\npdecimalsign{.}
\npproductsign{\cdot}

\usetikzlibrary{arrows,decorations.pathmorphing,positioning,fit,trees,shapes,shadows,automata,calc}

\begin{document}
\title[High-level Counterexamples for Probabilistic Automata]{High-level Counterexamples \\ for Probabilistic Automata}

\author[R. Wimmer]{Ralf Wimmer\rsuper a}
\author[N. Jansen]{Nils Jansen\rsuper b}
\author[E. \'Abrah\'am]{Erika \'Abrah\'am\rsuper c}
\author[J.-P. Katoen]{Joost-Pieter Katoen\rsuper d}

\address{{\lsuper a}Albert-Ludwigs-Universit\"at Freiburg, Germany}
\email{wimmer@informatik.uni-freiburg.de}
\address{{\lsuper{b,c,d}}RWTH Aachen University, Germany}
\email{\{nils.jansen $|$ abraham $|$ katoen\}@informatik.rwth-aachen.de}

\thanks{This work was partly supported by the German Research
  Council (DFG) as part of the Transregional Collaborative
  Research Center AVACS (SFB/TR~14), the 
  DFG project CEBug (AB 461/1-1), and the EU-FP7 IRSES project MEALS. Also funded by the
  Excellence Initiative of the German federal and state government.}

\begin{abstract}
  Providing compact and understandable counterexamples for violated
  system properties is an essential task in model checking. Existing
  works on counterexamples for probabilistic systems so far computed
  either a large set of system runs or a subset of the system's
  states, both of which are of limited use in manual debugging.  Many
  probabilistic systems are described in a guarded command language
  like the one used by the popular model checker \tool{PRISM}. In this
  paper we describe how a smallest possible subset of the commands can be
  identified which together make the system erroneous. We
  additionally show how the selected commands can be further
  simplified to obtain a well-understandable counterexample.
\end{abstract}

\maketitle


\section{Introduction}
\label{sec:introduction}

One of the main strengths---perhaps \emph{the} key feature---of model checking 
is its possibility to automatically generate a counterexample in case a model 
refutes a given property~\cite{Cla08}. Counterexamples provide essential 
diagnostic information for debugging purposes. They also play an important role 
in counterexample-guided abstraction refinement (CEGAR)~\cite{CGJLV00}, a 
successful technique in software verification. In this iterative 
abstraction-refinement process, abstractions that are too coarse are refined 
with the help of counterexamples. Single system runs---typically acquired during 
model checking---suffice as counterexamples for linear-time properties. For 
branching-time logics such as CTL and CTL$^*$, more general shapes are 
necessary, such as tree-like counterexamples~\cite{ClarkeJLV02}.

This paper focuses on counterexamples for \emph{probabilistic 
automata}~\cite{DBLP:journals/njc/SegalaL95}, shortly labeled transition systems 
in which transitions yield distributions over states (rather than just states). 
A violating behavior in this setting entails that the (maximal) probability that 
a certain property $\varphi$ holds, such as a deadlock state is reachable, is 
outside of some required bounds. For probabilistic reachability properties, it 
suffices to treat violations of upper bounds~\cite{HKD09}. Counterexamples 
consist of a finite set of finite runs that all satisfy the property $\varphi$ 
while their combined probability mass exceeds the required upper bound.  In 
contrast to some traditional model checking algorithms for LTL and CTL, 
probabilistic counterexamples are not obtained as a by-product of the 
verification process. Instead, dedicated counterexample generation algorithms 
are used.

In the last years, several approaches have been proposed for probabilistic 
counterexample generation. Enumerative approaches~\cite{AL10,HKD09,WBB09} 
generate a set of finite paths based on $k$ shortest path algorithms possibly 
enhanced with heuristic search and/or SAT-based techniques. Such counterexamples 
can be succinctly represented by, \eg regular expressions~\cite{HKD09} or in a 
hierarchical manner~\cite{ADR08,jansen-et-al-atva-2011} using a graph 
decomposition of the Markov chain into strongly connected components. An 
alternative is to use so-called \emph{critical 
sub-systems}~\cite{AL10,jansen-et-al-atva-2011}. The key idea of this approach 
is to obtain a---preferably small---connected fragment of the Markov chain that 
itself already violates the property at hand. This sub-system can thus be viewed 
as another representation of the set of runs that all satisfy the property 
$\varphi$ whose probability mass exceeds the required upper bound. 
In~\cite{wimmer-et-al-tacas-2012,wimmer-et-al-tcs-2014} we suggested to obtain 
\emph{minimal} critical sub-systems. Here, minimality refers to the state space 
size of the sub-Markov chain. Whereas~\cite{AL10,jansen-et-al-atva-2011} use 
heuristic approaches to construct small (but not necessarily minimal) critical 
sub-systems,~\cite{wimmer-et-al-tacas-2012,wimmer-et-al-tcs-2014} advocates the 
use of mixed integer linear programming (MILP)~\cite{Sch86}. The MILP-approach 
is applicable to $\omega$-regular properties (that include reachability) for 
both Markov chains and Markov decision processes 
(MDPs)~\cite{wimmer-et-al-tcs-2014}, which are a slightly variant of 
probabilistic automata.\footnote{Obtaining minimal critical sub-systems for MDPs 
is NP-complete~\cite{AL10}, as is solving MILP problems.} A more detailed overview 
of probabilistic counterexamples is given in \cite{abraham-et-al-sfm-2014}.

Despite the algorithmic differences, all counterexample generation algorithms 
published so far have one thing in common---they are all \emph{state} based. 
This means that path-based approaches yield paths in the Markov model, whereas 
the critical sub-system techniques obtain fragments of the Markov model. They do 
not obtain diagnostic information in terms of the modeling formalism in which 
these models are described. This seriously hampers the comprehensibility of 
counterexamples and is a significant obstacle in debugging the system model 
(description). In addition, as most practical systems consist of various 
components running concurrently, counterexamples in terms of the underlying 
(potentially huge) state space are often too large to be used effectively. 
Although critical sub-systems often are orders of magnitude smaller than the 
original system, they may still be very large, rendering manual debugging 
practically impossible.

To overcome these deficiencies, this paper focuses on obtaining 
\emph{counterexamples at the level of the modeling formalism.} The basic idea is 
to determine a minimal critical model description that acts as a 
counterexample---preferably in a fully automated manner. Intuitively speaking, 
our approach determines the fragments of the model description that are relevant 
for the violation of the property at hand. Having a human-readable specification 
language, it seems natural that a user should be pointed to the part of the 
system description which causes the error. This is exactly what our approach for 
\emph{high-level counterexamples} attempts to accomplish.

We assume Markov models are described using a stochastic version of Alur and 
Henzinger's reactive modules~\cite{DBLP:journals/fmsd/AlurH99b}. This is the 
modeling formalism adopted by the popular probabilistic model checker 
\tool{PRISM}~\cite{KNP11}. In this setting, a probabilistic 
automaton~\cite{DBLP:journals/njc/SegalaL95} model is typically specified as a 
parallel composition of modules. The behavior of a single module is described 
using a probabilistic extension~\cite{HSM97} of Dijkstra's \emph{guarded command 
language}~\cite{Dij75}. Modules communicate by shared variables or 
synchronization on common actions. Our approach however is also applicable to 
other modeling formalisms for probabilistic automata such as the process 
algebraic approach in~\cite{DBLP:journals/tcs/KatoenPST12} by using the 
similarities between linearised process descriptions and the guarded command 
language.

This paper considers the problem of determining a---preferably small---set of 
guarded commands that together induce a critical sub-system. In order to correct 
the system description, at least one of the returned guarded commands has to be 
changed. We show how to simplify the commands by removing command branches which 
are not necessary to obtain a counterexample. We present this as a special case 
of a method where the number of different transition labels for a probabilistic 
automaton is minimized. This offers great flexibility in terms of human-readable 
counterexamples. We show that obtaining a \emph{minimal} command set that acts 
as a counterexample is NP-complete and advocate the usage of MILP techniques to 
obtain such a smallest critical label set. Besides the theoretical principles of 
our technique, we illustrate its practical feasibility by showing the results of 
applying a prototypical implementation to various examples from the \tool{PRISM} 
benchmark suite.

\medskip

\noindent\textit{Structure of the paper.}  
The first section brief{}ly reviews the necessary foundations. 
Section~\ref{sec:minimization} presents the theoretical framework to obtain minimal counterexamples.
Section~\ref{sec:simplification} introduces several simplification steps for reactive modules.
After presenting some experimental results in Section~\ref{sec:experiments}, we conclude the paper in
Section~\ref{sec:conclusion}.

\medskip

\noindent 
This paper is an extended version of~\cite{wimmer-et-al-qest-2013}. 
The extensions include (1) a more general labeling (branches instead of transitions) enabling more 
simplification steps, (2) minimization of variable values and intervals, and (3) detailed correctness 
proofs for the MILP formulation.


\section{Foundations}
\label{sec:foundations}

\noindent Let $S$ be a countable set. A \emph{sub-distribution} on $S$
is a function $\mu:S\to \Irational$ such that $0<\sum_{s\in S}\mu(s) \leq
1$ with support $\supp(\mu)=\{s\in S\, | \, \mu(s)>0\}$. We use the
notation $\mu(S')=\sum_{s\in S'}\mu(s)$ for $S'\subseteq S$. A
sub-distribution with $\mu(S)=1$ is called a \emph{probability
  distribution}. We denote the set of all probability distributions on
$S$ by $\Distr(S)$ and analogously the set of sub-distributions by
$\SubDistr(S)$.

\subsection{Probabilistic Automata}
\label{subsec:pa}

\begin{defi}[Probabilistic automaton]
  \label{def:pa}
  A \emph{probabilistic automaton (PA)} is a tuple
  $\PA=(S,\sinit,\Act,P)$ such that $S$ is a finite set of states,
  $\sinit\in S$ is an initial state, $\Act$ is a finite set of
  actions, and $P: S\to (2^{\Act\times\Distr(S)}\setminus\{\emptyset\})$ is a probabilistic
  transition relation such that $P(s)$ is finite for all $s\in S$.
\end{defi}

For notational convenience, we use
all notations defined for (sub-)distributions also for
action-distribution pairs with the natural meaning, \eg
$\supp(\edge)=\supp(\mu)$ and $\edge(s)=\mu(s)$ for
$\edge=(\act,\mu)\in \Act\times\Distr(S)$.
We call $\edge\in P(s)$ a \emph{transition}, while a
tuple $(s,\edge,s')$ with $\edge\in P(s)$ and $\edge(s')>0$ is called a
\emph{branch} of the transition. 

For a state $s\in S$, a successor state is determined as follows: A
transition $\edge\in P(s)$ is chosen non-deterministically. Then,
$s'\in\supp(\edge)$ is determined probabilistically
according to the distribution in $\edge$. This process can be repeated
infinitely often starting with the initial state $\sinit$. To prevent
deadlocks we assume $P(s)\neq\emptyset$ for all $s\in S$.

An \emph{infinite path} of a PA $\PA$ is an infinite sequence
$s_0\edge_0s_1\edge_1s_2\ldots$ with $s_i\in S$,
$\edge_i\in P(s_i)$ and
$s_{i+1}\in\supp(\edge_i)$ for all $i\geq 0$. A
\emph{finite path} $\pi$ of $\PA$ is a finite prefix
$s_0\edge_0s_1\edge_1\ldots s_n$ of an infinite path
of $\PA$ with $\last(\pi)=s_n$. The
set of all finite paths of $\PA$ is $\pathsFin_\PA$.

\smallskip

A \emph{sub-PA} is like a PA, but it allows sub-distributions instead
of distributions in the definition of $P$.

\begin{defi}[Sub-PA]
  \label{def:sub_pa}
  A \emph{sub-PA} is a tuple $\PA=(S,\sinit,\Act,P)$ with $S$,
  $\sinit$, and $\Act$ as in Definition~\ref{def:pa} and $P: S\to
  2^{\Act\times\SubDistr(S)}$ is a probabilistic transition relation such
  that $P(s)$ is finite for all $s\in S$.
\end{defi}

A sub-PA $\PA=(S,\sinit,\Act,P)$ can be transformed into a PA as
follows: We add a new state $s_{\bot}\not\in S$ and a new action
$\tau\notin\Act$, extend all sub-distributions into probability
distributions by defining $\mu(s_{\bot}) = 1 - \mu(S)$ for
each $s\in S$ and $(\act,\mu)\in P(s)$, and set
$P(s)=\{(\tau,\mu)\in \{ \tau \} \times\Distr(S\cup\{s_{\bot}\})\ |
\ \mu(s_{\bot})=1\}$ for each $s\in \{s_{\bot}\}\cup\{s'\in S\, | \,
P(s')=\emptyset\}$. 
This allows for applying all methods we use for PAs also for sub-PAs.

\begin{defi}[Subsystem]
  \label{def:subsystem}
  A sub-PA $\PA'=(S',\sinit',\Act',P')$ is a \emph{subsystem} of a sub-PA
  $\PA=(S,\sinit,\Act,P)$, written $\PA'\sqsubseteq \PA$, iff $S'\subseteq S$,
  $\sinit' = \sinit\in S'$, $\Act'\subseteq\Act$ and for all $s\in S'$ there
  is an injective function $f:P'(s)\rightarrow P(s)$ such that for all
  $(\act',\mu')\in P'(s)$ with $f((\act',\mu'))=(\act,\mu)$ we have that
  $\act'=\act$ and for all $s'\in S'$ either $\mu'(s')=0$ or
  $\mu'(s')=\mu(s')$.
\end{defi}

In this paper we are interested in
\emph{probabilistic reachability properties}: Is the probability to
reach a set $T\subseteq S$ of target states from $\sinit$ at most a
given bound $\lambda\in[0,1]\subseteq\mathbb{R}$? This 
property is denoted by
$\pctlProb{\leq\lambda}{\finally{T}}$. Note that checking arbitrary
$\omega$-regular properties can be reduced to checking reachability
properties, see~\cite[Chapter 10.3]{BK08}. To define a suitable probability measure on PAs, the
nondeterminism has to be resolved. This is done by an oracle 
called \emph{scheduler}.

\begin{defi}[Scheduler]
  \label{def:scheduler}
  A memoryless deterministic \emph{scheduler}\footnote{Note that
    schedulers in their full generality are functions mapping finite
    paths of the PA to distributions over the outgoing transitions of
    each state.  For unbounded probabilistic reachability properties
    memoryless deterministic schedulers suffice to obtain maximal (and
    minimal) probabilities~\cite[Lemma~10.102]{BK08}.} for a sub-PA $\PA=(S,\sinit,$ $\Act,P)$
  is a partial function $\sigma:S\partialto\Act\times\SubDistr(S)$ with
  $\sigma(s)\in P(s)$ for all $s\in\dom(\sigma)$.  We use
  $\Sched_{\PA}$ to denote the set of all memoryless deterministic
  schedulers of $\PA$. $\Sched^+_{\PA}\subseteq \Sched_{\PA}$ is the
  set of all schedulers that are total functions
  $\sigma:S\to\Act\times\SubDistr(S)$. Such schedulers are called
  \emph{deadlock-free}.
\end{defi}

As a scheduler resolves the nondeterminism for a PA, this induces a
fully probabilistic model, for which a standard probability measure
can be defined. We refer to~\cite[Chapter 10.1]{BK08} for more details
on schedulers and measure theory.

\begin{defi}[Sub-PA induced by scheduler]\label{def:induced_pa}
  For a sub-PA $\PA=(S,\sinit,\Act,P)$ and a scheduler
  $\sigma\in\Sched_{\PA}$, the \emph{sub-PA induced by $\PA$ and
    $\sigma$} is given by
  $\PA^\sigma=(S^\sigma,\sinit,\Act^\sigma,P^\sigma)$ with
  $S^\sigma=S$, $\Act^\sigma=\{\act\in\Act\mid \exists s\in
  S\suchthat\exists\mu\in\SubDistr(S)\suchthat\sigma(s)=(\act,\mu)\}$, and $\
  P^\sigma(s)=\{ \sigma(s) \}$ for all $s\in \dom(\sigma)$ and $P^\sigma(s)=\emptyset$
  for all $s\in S\setminus\dom(\sigma)$.
\end{defi}
Note that in $\PA^\sigma$, $|P^{\sigma}(s)|\leq 1$ holds for all $s\in S$. In
fact, the actions could be removed, which would yield a
\emph{discrete-time Markov chain}. For details, we again refer
to~\cite{BK08}.

For a sub-PA $\PA$ with a fixed scheduler $\sigma$, the probability
$\prob_{\PA^{\sigma}}(\sinit,\finally{T})$ can now be computed by
solving a linear equation system. The property
$\pctlProb{\leq\lambda}{\finally{T}}$ is satisfied by a sub-PA $\PA$
if $\prob_{\PA^{\sigma}}(\sinit,\finally{T})\leq\lambda$ holds for all
schedulers $\sigma$ for $\PA$. To check this, it suffices to compute
the maximal probability to reach $T$ from
$\sinit$ over all schedulers, which we denote by $\probmax_\PA(\sinit,\finally{T})$. This
probability is given by the unique solution of the following equation
system:
\begin{eqnarray}
\label{eq:reachprob}
\probmax_\PA(s,\finally{T})=
	\begin{cases}
		1 \quad \text{ if } s\in T, \\
		0 \quad \text{ if } T \text{ is unreachable from $s$ under all schedulers},\\
		\displaystyle \max\limits_{\edge\in P(s)} \sum\limits_{s'\in S}\edge(s')\cdot\probmax_\PA(s',\finally{T}) \quad \text{ otherwise.}
	\end{cases}
\end{eqnarray}
It can be solved by either rewriting it into a linear program, by applying a
technique called value iteration, or by iterating over the possible
schedulers (policy iteration) (see, \eg \cite[Chapter 10.6]{BK08}).
A memoryless deterministic scheduler can be obtained easily from the solution
of the equation system (cf.~\cite[Lemma 10.102]{BK08}).

\subsection{\tool{PRISM}'s Guarded Command Language}
\label{subsec:gcl}

For a set $\Vars$ of bounded integer variables, let $\Assignments{\Vars}$ denote
the set of all variable assignments, \ie of functions $\nu:\Vars\to\mathbb{Z}$
such that $\nu(\var)\in\dom(\var)$ for all $\var\in\Vars$. We assume that the domains
of all variables are finite.
\begin{defi}[Model, module, command]
  \label{def:prism_model}
  A \emph{model} is a tuple $(\Vars,\sinit,\{M_1,\ldots,M_k\})$ where
  $\Vars$ is a finite set of Boolean variables, 
  $\sinit \in \Assignments{\Vars}$ an initial assignment, and 
  $\{M_1,\ldots,M_k\}$ a finite set of modules.

  A \emph{module} is a tuple $M_i = (\Vars_i,\Act_i,C_i)$ with
  $\Vars_i\subseteq\Vars$ a set of variables such that
  $\Vars_i\cap\Vars_j=\emptyset$ for $i\not= j$, $\Act_i$ a finite set
  of synchronizing actions, and $C_i$ a finite set of commands. The
  action $\tau$ with $\tau\not\in\bigcup_{i=1}^k\Act_i$ denotes the
  internal non-synchronizing action. A \emph{command} $c\in C_i$ has the form
  \[
    c\ =\ [\act]\ g\to\ p_1: f_1 + \ldots + p_n: f_n
  \] 
  with $\act\in\Act_i\dcup\{\tau\}$, $g$ a Boolean predicate
    (``guard'') over the variables in $\Vars$, $p_j\in[0,1]$ a
    rational number with $\sum_{j=1}^n p_j = 1$, and $f_j:
    \Assignments{\Vars}\to\Assignments{\Vars_i}$ being a variable
    update function. We refer to the action $\act$ of command $c$ by
    $\actionof{c}$.
\end{defi}

Note that each module may only change the values of its own variables
while their new values may depend on variables of other modules. Each
model with several modules is equivalent to a model with a single
module, which is obtained by computing the parallel composition of
these modules. We give a short intuition on how this composition is
built. For more details we refer to the documentation of \tool{PRISM}.

Assume two modules $M_1=(\Vars_1,\Act_1,C_1)$ and
$M_2=(\Vars_2,\Act_2,C_2)$ with $\Vars_1\cap\Vars_2=\emptyset$. We 
first define the composition $c\otimes c'$ of two commands $c$ and $c'$:
For $c = [\act]\ g\to\ p_1: f_1 + \ldots + p_n:
f_n\in C_1$ and $c' = [\act]\ g'\to\ p_1': f_1' + \ldots + p_m':
f_m'\in C_2$ we have:
\[
  c\otimes c'\ =  \ [\act]\ g\land g'\quad\to\quad \sum_{i=1}^n\sum_{j=1}^m p_i\cdot p_j': f_i\otimes f_j'\ .
\]
Here, for $f_i:\Assignments{\Vars}\to\Assignments{\Vars_1}$ and
$f_j':\Assignments{\Vars}\to\Assignments{\Vars_2}$ we define
$f_i\otimes f_j':
\Assignments{\Vars}\to\Assignments{\Vars_1\cup\Vars_2}$ such that
for all $\nu\in\Assignments{\Vars}$ we have that 
$(f_i\otimes f_j')(\nu)(\var)$ equals $f_i(\nu)(\var)$ for each $\var\in \Vars_1$
and $f_j'(\nu)(\var)$ for each $\var\in \Vars_2$.

Using this, the \emph{parallel composition} $M=M_1\mathbin{\|}M_2=(\Vars,\Act,C)$
is given by $\Vars = \Vars_1 \cup \Vars_2$, $\Act = \Act_1 \cup \Act_2$, and
\[
\begin{array}{llcc@{|}ll}
  C & = & 
    \{&
    c\ &\, c\in C_1 \cup C_2 \wedge \actionof{c} \in \{\tau\}\cup(\Act_1\setminus\Act_2) \cup (\Act_2\setminus\Act_1) & \} \cup {} \\[1ex]
&&    \{& c\otimes c'\, & \,
      c\in C_1 \wedge c'\in C_2 \wedge \actionof{c}=\actionof{c'}\in \Act_1\cap\Act_2 & \}\ .
\end{array}
\]

Intuitively, commands labeled with non-synchronizing actions are executed on
their own, while for synchronizing actions a command from each synchronizing
module is executed simultaneously. Note that if a module has an action in its
synchronizing action set but no commands labeled with this action, this
module will block the execution of commands with this action in the composition.
This is considered to be a modeling error and the corresponding commands are
ignored.

The PA-semantics of a model is as follows. Assume a model
$(\Vars,\sinit,\{M\})$ with a single module $M=(\Vars,\Act,C)$ which
will not be subject to parallel composition any more and $\Vars=\{\var_1,\ldots,\var_m\}$. The
\emph{state space} $S$ of the corresponding PA $\PA=(S,\sinit,\Act,P)$
is given by the set of all possible variable assignments
$\Assignments{\Vars}$, \ie a state $s$ is a vector
$(v_1,\ldots,v_m)$ with $v_i$ being a value of the variable
$\var_i\in\Vars$. To construct the transitions,
we observe that the guard $g$ of each command
\[
  c\ =\ [\alpha]\ g\to\ p_1: f_1 + \ldots + p_n: f_n\ \in C
\]
defines a subset of the state space $S_c\subseteq
\Assignments{\Vars}$ with $s\in S_c$ iff $s$ satisfies $g$. 
For each state $s\in S_c$ we define
a probability distribution $\mu_{c,s}:\Assignments{\Vars}\to[0,1]$
with
\[
  \mu_{c,s}(s')=\sum_{\{1\leq i\leq n\mid f_i(s)=s'\}} p_i
\]
for each $s'\in\Assignments{\Vars}$. The probabilistic transition
relation $P:\Assignments{\Vars}\to 2^{\Act\times\Distr(\Assignments{\Vars})}$ is given by
$P(s)=\{(\act,\mu_{c,s})\mid c\in C\wedge \actionof{c}=\act\wedge s\in S_c\}$ for all
$s\in\Assignments{\Vars}$.

\begin{exa}
  \label{ex:coin}
  We consider the shared coin protocol of a
  randomized consensus algorithm~\cite{consensus}. The protocol returns
  a preference between two choices  with a
  certain probability. A shared integer variable\footnote{Our simplified definition
  of a model does not support variables which are written by more than one module.
  The language actually implemented by PRISM, however, allows such variables with the
  restriction that they may be written only by non-synchronizing commands in 
  order to avoid writing conflicts.} \texttt{c}
  is incremented or decremented by each process depending on the
  internal result of a coin flipping. If the value of \texttt{c} becomes
  lower than a threshold \texttt{left} or higher than a threshold \texttt{right},
  the result is \texttt{heads} or
  \texttt{tails}, respectively.

  The protocol, which is the same for each participating
  process, has the following local variables: \texttt{coin} which
  is either $0$ or $1$, \texttt{flip} which is \texttt{true} iff the
  coin shall be flipped, \texttt{flipped} which is \texttt{true} iff
  the coin has already been flipped, \texttt{check} which is
  \texttt{true} iff the value of \texttt{c} shall be checked.
  Initially, \texttt{c} has a value between \texttt{left} and
  \texttt{right}, \texttt{flip} is true, and \texttt{flipped} and
  \texttt{check} are false. Consider a simplified version
  of the original \tool{PRISM} code:
  
  \begin{small}\begin{align}
      [\tau] &\ \texttt{flip} & \rightarrow\  & 0.5:\texttt{coin=0}  \,\&\,  \texttt{flip=false}  \,\&\, \texttt{flipped=true}\notag\\& & &+ 0.5:\texttt{coin=1}  \,\&\,  \texttt{flip=false}  \,\&\,  \texttt{flipped=true}\label{coin_flip}\\
      [\tau] &\ \texttt{flipped}  \,\&\, \texttt{coin=0} \,\&\, \texttt{c$\leq$right} & \rightarrow\ & 1:\ \texttt{c=c-1}  \,\&\, \texttt{flipped=false}  \,\&\, \texttt{check=true}\label{coin_decrement}\\
      [\tau] &\ \texttt{flipped}  \,\&\, \texttt{coin=1} \,\&\, \texttt{left$\leq$c} & \rightarrow\ & 1:\ \texttt{c=c+1}  \,\&\, \texttt{flipped=false}  \,\&\, \texttt{check=true}\label{coin_increment}\\
      [\tau] &\ \texttt{c<left} & \rightarrow\ & 1:\ \texttt{heads=true}\label{coin_heads}\\
      [\tau] &\ \texttt{c>right} & \rightarrow\ & 1:\ \texttt{tails=true}\label{coin_tails}\\
      [\tau] &\ \texttt{check}  \,\&\, \texttt{c$\leq$right}  \,\&\, \texttt{c$\geq$left} & \rightarrow\ & 1:\ \texttt{check=false}  \,\&\, \texttt{flip=true} \label{coin_flip_again}
    \end{align}\end{small}

\noindent   Command~\ref{coin_flip} sets \texttt{coin} to $0$ or $1$, each with
  probability $0.5$. Commands~\ref{coin_decrement}
  and~\ref{coin_increment} increment or decrement the shared counter
  \texttt{c} depending on the value of
  \texttt{coin}. Commands~\ref{coin_heads} and~\ref{coin_tails} check
  whether the value of \texttt{c} is above or below the boundaries
  \texttt{left} and \texttt{right} and return \texttt{heads} or
  \texttt{tails}, respectively. If no boundary is violated,
  Command~\ref{coin_flip_again} sets \texttt{flip} to \texttt{true}
  which enables Command~\ref{coin_flip} again.
\end{exa}

\subsection{Mixed Integer Programming}
\label{subsec:milp}

\noindent A \emph{mixed integer linear program} optimizes a linear objective function
under a condition specified by a conjunction of linear inequalities. A
subset of the variables in the inequalities is restricted to take
only integer values, which makes solving MILPs
NP-hard~\cite[Problem MP1]{GJ79}.

\begin{defi}[Mixed integer linear program]
  \label{def:milp}
  Let $A\in\mathbb{Q}^{m\times n}$, $B\in\mathbb{Q}^{m\times k}$,
  $b\in\mathbb{Q}^m$, $c\in\mathbb{Q}^n$, and $d\in\mathbb{Q}^k$.
  A \emph{mixed integer linear program} (MILP) consists in computing
  $\min c^Tx + d^Ty$ such that $Ax + By \leq b$ and $x\in\mathbb{R}^n,\ y\in\mathbb{Z}^k$.
\end{defi}

MILPs are typically solved by a combination of a branch-and-bound algorithm and
the generation of so-called cutting planes. These algorithms heavily rely on the
fact that relaxations of MILPs which result from removing the integrality
constraints can be efficiently solved. MILPs are widely used in operations
research, hardware-software co-design, and numerous other applications. Efficient
open source as well as commercial implementations are available like \tool{Scip}~\cite{Ach09},
and \tool{Gurobi}~\cite{gurobi}. We refer to the textbook
\cite{Sch86} for more information on solving MILPs.


\section{Computing Counterexamples}
\label{sec:minimization}

In this section we show how to compute a smallest critical command set of
a given model, \ie,
a smallest subset of the model's commands which lead to an erroneous
system independent of the other commands. For this, we introduce a 
generalization of this problem, namely smallest
critical label sets, state the complexity, and specify an
MILP formulation to solve this problem.

\subsection{Smallest Critical Label Sets}
\label{subsec:scl}

Let $\PA=(S,\sinit,\Act,P)$ be a PA, $T\subseteq S$, and $\Lab$ a
finite set of labels. Assume furthermore a partial labeling function
$\LT:S\times\Act\times\Distr(S)\times S\partialto 2^{\Lab}$ such that
$\LT(s,\edge,s')$ is defined iff $\edge\in P(s)$ and
$s'\in\supp(\edge)$.

Let $\Lab'\subseteq\Lab$ be a subset of the labels. The \emph{sub-PA
induced by $\Lab'$} is $\PAind=(S,\sinit,\Act,P_{|\Lab'})$ such that 
for all $s\in S$ we have
\begin{eqnarray*}
 P_{|\Lab'}(s) &=& \{
 (\act,\mu_{|\Lab'})
\,\mid\,
(\act,\mu)\in P(s)\wedge \exists s'\in S \suchthat L(s,\act,\mu,s')\subseteq\Lab'\}\, ,
\end{eqnarray*}
where $\mu_{|\Lab'}\in\SubDistr(S)$ with $\mu_{|\Lab'}(s')=\mu(s')$ if
$\LT(s,\act,\mu,s')\subseteq\Lab'$ and $\mu_{|\Lab'}(s')=0$ otherwise
for each $s'\in S$.
Thus in $\PAind$ all branches have been removed whose labeling is not
a subset of $\Lab'$.

\begin{defi}[Smallest critical label set (SCL) problem]
  \label{def:scl}
  Let $\PA$, $T$, $\Lab$, and $\LT$ be defined as above
  and $\pctlProb{\leq\lambda}{\finally{T}}$ be a reachability property
  that is violated by $\sinit$ in $\PA$. A label set $\Lab'\subseteq\Lab$ 
  and its induced sub-PA $\PAind = (S,\sinit,\Act,P')$ 
  are called \emph{critical} if 
  $\probmax_{\PAind}(\sinit,\finally{T})>\lambda$.

  Given a weight function $w:\Lab\to\mathbb{R}^{\geq 0}$, 
  the \emph{smallest critical label set} (\emph{SCL}) \emph{problem} is to 
  determine a critical subset $\Lab'\subseteq\Lab$ such that
  $w(\Lab')= \sum_{\ell\in\Lab'}w(\ell)$ is minimal 
  among all critical subsets of $\Lab$.
\end{defi}

\begin{thm}
  \label{th:np_complete}
  To decide whether there is a critical label set $\Lab'\subseteq\Lab $ 
  with $w(\Lab') \leq k$ for a given integer $k\geq 0$ is NP-complete.
\end{thm}

\noindent The proof of this theorem is a reduction from exact 3-cover
(X3C)~\cite{GJ79}, similar to a proof in~\cite{CV10}. \IfLongVersion{For the aid
of the reviewers, we give the proof in Appendix~\ref{app:proof_np}.}

The concept of smallest critical label sets gives us a flexible
description of counterexamples being minimal with respect to
different quantities. We will now list different kinds of 
counterexamples that can be computed using an SCL.

\paragraph{\bfseries Commands} In order to minimize the number of
commands that together induce an erroneous system, \ie form a \emph{critical command set}, let
$\PA=(S,\sinit,\Act,P)$ be a PA generated by modules
$M_i=(\Vars_i,\Act_i,C_i)$, $i=1,{\ldots},k$. For each module $M_i$
and each command $c\in C_i$ we introduce a unique label\footnote{In
the following we write short $\ell_c$ instead of $\ell_{c,i}$ if the
index $i$ is clear from the context.} $\ell_{c,i}$ with weight $1$ and
define the labeling function 
$\LT:S\times\Act\times\Distr(S)\times S\partialto 2^{\Lab}$ such that the
labels in $\LT(s, \edge, s')$ correspond to the set of commands which
together generate this transition $\eta\in P(s)$.\footnote{If
several command sets generate the same transition, we introduce copies of
the transition.} Note that in case of synchronization several
commands together create a certain transition. An SCL then corresponds 
to a \emph{smallest critical command set}.

\paragraph{\bfseries Modules} We can also minimize the number of
modules involved in a counterexample by using the same label for all
commands in a module. Often systems consist of a number of copies of
the same module, containing the same commands, only with the local variables
renamed, plus a few extra modules. Consider for example a
wireless network: $n$ nodes want to transmit messages using a protocol
for medium access control~\cite{wlan}. All nodes run the same
protocol. Additionally, there may be a module describing the
channel. When fixing an erroneous system, one wants to preserve the
identical structure of the nodes. Therefore the selected commands
should contain the same subset of commands from all identical
modules. This can be obtained by assigning the same label to all
corresponding commands from the symmetric modules and using the number
of symmetric modules as its weight.

\paragraph{\bfseries Deletion of unnecessary branches}

The SCL problem can also be used to simplify commands.
For this we 
identify a smallest set of command branches
that need to be preserved, such that the induced sub-PA still violates
the property under consideration.
The resulting command branches can be removed, still yielding an erroneous system. 
Given a command $c_i$ of the form
$
  [\act]\ g \to p_1: f_1 + p_2: f_2 + \cdots + p_{n}: f_{n},
$
we assign to each command branch $p_j: f_j$ a unique label $\ell_{i,j}$ with weight $1$. 
Let $\Lab$ be the set of all such labels. 
When the parallel composition of modules is computed, see Section~\ref{subsec:gcl}, we build the union of
the labelings of the synchronizing command branches being executed together. When
computing the corresponding PA $\PA$, we transfer this labeling to the transition branches
of $\PA$: We define the labeling function
$\LT$ such that $\LT(s,\edge,s')$ contains the labels of all command branches
that are involved in generating the branch
from $s$ to $s'$ via the transition $\edge$.

\paragraph{\bfseries States} The state-minimal critical subsystems as
introduced in~\cite{wimmer-et-al-tacas-2012} can be also obtained as
special case of smallest critical label sets: For each state $s\in S$
introduce a label $\ell_s$ with weight $1$ and set
$\LT(s,\edge,s')=\{\ell_{s'}\}$ for all $s\in S$, $\edge\in P(s)$ and
$s'\in\supp(\edge)$. Then a smallest critical label set
$\Lab'\subseteq\Lab=\{\ell_s\,|\,s\in S\}$ induces a state-minimal critical subsystem.

\paragraph{\bfseries Variable domains} Smallest critical label sets can also
be used to reduce the domains of the variables in the \tool{PRISM} program.
Let $\Vars$ be the set of variables of a \tool{PRISM} program and
$\PA=(S,\sinit,\Act,P)$ the corresponding PA. Note that each state
$s\in S$ corresponds to an assignment of the variables in $\Vars$.
For a variable $\var\in\Vars$ we denote by $s(\var)$ the value of $\var$ in 
state $s$. Let $\Lab = \{ \ell_{\var,v}\,|\,\var\in\Vars\land v\in\dom(\var)\}$ be
the set of labels, each with weight 1.
We define the labeling of transition branches by corresponding variable values as
$\LT(s,\edge,s')=\{\ell_{\var,v} \,|\, \var\in\Vars\wedge s'(\var)=v\}$.
A smallest critical labeling induces a critical subsystem with a minimum number of
variable values. If we restrict the variable domains to these values,
we still obtain an erroneous system.

\paragraph{\bfseries Variable intervals} The previous reduction technique
removes a maximum number of values from the variables' domains. Originally
the domains are intervals in $\mathbb{Z}$. Minimization, however,
yields sets that are in general not intervals anymore. We
can also minimize the size of the intervals instead. To do so we
need to impose further constraints on the valid label sets $\Lab'$. Details
will be presented in Section~\ref{subsec:interval_reduction}.

\begin{rem}
 \label{rem:size_of_label_set}
 The various applications require label sets of very different sizes.
 For the minimization of commands, modules, and branches the number of
 labels is linear in the size of the model description. In contrast, the number of different
 labels for state minimization is linear in the number of (reachable) states
 of the described PA, which can be exponential in the size of its description.
 The same holds for the minimization of variable domains and intervals.
\end{rem}

\subsection{Computing Smallest Critical Label Sets}

\noindent We now explain how smallest critical label sets can be
computed. First, the notions of \emph{relevant} and \emph{problematic}
states are introduced. Intuitively, state $s$ is relevant if it is
on a path from the initial state to a target state. A relevant state $s$ is
problematic, if there exists a deadlock-free scheduler under which no target state 
is reachable from $s$.

\begin{defi}[Relevant and problematic states and transitions]
  \label{def:relevant_states}
  Let $\PA$, $T$, and $\LT$ be as above. The
  \emph{relevant states of $\PA$ for $T$} are given by $\relstates_T =
  \{s\in S\,|\,\exists \sigma\in\Sched_\PA\suchthat$ $
  \prob^{\sigma}_\PA(\sinit,\finally{\{s\}})>0\land \prob^{\sigma}_\PA(s,\finally{T}) > 0\}$.  
  A label $\ell$ is \emph{relevant} for $T$ if there are $s\in\relstates_T$, 
  $\edge\in P(s)$, and $s'\in\supp(\edge)\cap\relstates_T$ such that 
  $\ell\in\LT(s,\edge,s')$.

  The states in 
  $\probstates_T = \{s\in \relstates_T\,|\,\exists\sigma\in\Sched_\PA^+\suchthat\prob^{\sigma}_\PA(s,\finally{T}) = 0\}$
  are \emph{problematic states} and the set
  $\probtrans_T = \bigl\{ (s,\edge)\in \probstates_T\times\Act\times\Distr(S)\,\big|\,\edge\in P(s)\land\supp(\edge)\subseteq\probstates_T\bigr\}$ 
  are \emph{problematic state-transition-pairs} regarding $T$.
\end{defi}

States that are not relevant can be removed from the PA together with
all their incident branches without changing the probability of reaching
$T$ from $\sinit$. Additionally, all labels that do not occur in the
relevant part of the PA can be deleted. We therefore assume that the
(sub-)PA under consideration contains only states and labels that are
relevant for $T$.  Note that the relevant states and labels can
be computed in linear time using graph algorithms~\cite{BdA95}.
In our computation, we need to ensure that from each problematic state
a non-problematic state is reachable under the selected scheduler,
otherwise the probability of problematic states is not well-defined by 
the constraints as in~\cite{atr88}.
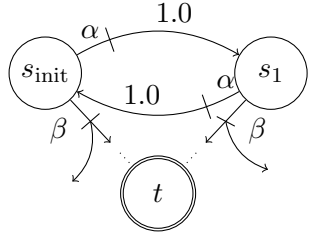
\begin{figure}[tb]
  \centering
  \begin{tikzpicture}[state/.append style={node distance=3cm,inner sep=1mm},scale=0.8]
    \node[state] (s0) {$\sinit$};
    \node[state] (s1) [right of=s0] {$s_1$};
    \node[state,accepting] (sn) at ($0.5*(s0)+0.5*(s1)+(0,-2)$) {$t$};
    
    \path[->] (s0) edge[bend left] node[above,pos=0.08]{$\act$} node[pos=0.2] (p1){} node[above,pos=0.6]{1.0} (s1);
    \draw[-] ($(p1)+(-0.7mm,2mm)$) -- ($(p1)+(0.7mm,-2mm)$);
    
    \path[->] (s1) edge[bend left] node[above,pos=0.08]{$\act$} node[pos=0.2] (p2){} node[above,pos=0.6]{1.0} (s0);
    \draw[-] ($(p2)+(-0.7mm,2mm)$) -- ($(p2)+(0.7mm,-2mm)$);
        
    \path (s0) -- node[pos=0.8](tip){} node[below left,pos=0.12]{$\beta$} (sn);
    \path[->] (s0) edge node[pos=0.5](tipa){} (tip);
    \draw[-] ($(tipa)+(-1.5mm,-1.3mm)$) -- ($(tipa)+(1.5mm,1.3mm)$);
    \path[->] (tipa.center) edge[bend left] ($(tipa) + (-3mm,-1cm)$);
    \draw[-] (tip.center) edge[dotted] (sn);
        
    \path (s1) -- node[pos=0.8](tipp){} node[below right,pos=0.12]{$\beta$} (sn);
    \path[->] (s1) edge node[pos=0.5](tipb){} (tipp);
    \draw[-] ($(tipb)+(-1.5mm,0.8mm)$) -- ($(tipb)+(1.5mm,-0.8mm)$);
    \path[->] (tipb.center) edge[bend right] ($(tipb) + (7mm,-8mm)$);
    \draw[-] (tipp.center) edge[dotted] (sn);
  \end{tikzpicture}
  \caption{Example for problematic states}
  \label{fig:problematic}
\end{figure}

\begin{exa}
  \label{ex:problematic}
  The PA in Figure~\ref{fig:problematic} illustrates the issues with problematic states.
  Assume $t$ is a target state. States $\sinit$ and $s_1$ are both problematic since 
  the scheduler which selects $\act$ in both $\sinit$ and $s_1$ prevents reaching a 
  target state, but all other schedulers do not. We cannot remove the outgoing transitions
  belonging to action $\act$ in a preprocessing step since a scheduler may choose $\act$ 
  in one state and $\beta$ in the other one. However, if a scheduler chooses $\act$ in
  both states, we obtain according to Equation~\eqref{eq:reachprob} the following equation system for model checking:
  \[
    p_{\sinit} =  1.0\cdot p_{s_1} \qquad
    p_{s_1} = 1.0\cdot p_{\sinit}
  \]
  For all $\kappa\in[0,1]$ we obtain a
  solution by setting $p_{\sinit}=p_{s_1}=\kappa$, although the target state $t$ is 
  not reachable at all.
\end{exa}

We solve this problem by attaching a value $r_s$ to each problematic
state $s\in\probstates_T$ and encoding that a transition of $s$ may be
selected only if it has at least one successor state $s'$ with a value
$r_{s'}>r_s$ attached to it. Since the sub-PA is finite, this
requirement assures by induction that there is a (loop-free)
\emph{increasing path} from $s$ to a non-problematic or deadlock
state, along which the values attached to the states are strictly
increasing. This implies that the probability of each loop visiting
problematic states only is always less than one.

To encode the computation of smallest critical label sets as MILP, 
we use the following variables with values assigned as described:
\begin{itemize}
\item for each $\ell\in\Lab$ a variable $x_\ell\in\Iint$ which is 1 
  iff $\ell$ is part of the critical label set;
\item for each state $s\in S\setminus T$ and each transition $\edge\in P(s)$ a variable
  $\sigma_{s,\edge}\in\Iint$ which is $1$ iff $\edge$ is chosen in $s$ by the scheduler;
  the scheduler is free not to choose any transition;
\item for each branch $(s,\edge,s')$ with $s\in S$, $\edge\in P(s)$ and $s'\in\supp(\edge)$
  a variable $p_{s,\edge,s'}\in\Ireal$ which is $0$ if not all labels in $\LT(s,\edge,s')$ 
  are contained in $\Lab'$, and at most $\edge(s')$ otherwise;
\item for each state $s\in S$ a variable $p_s\in\Ireal$ whose value is at most the probability to 
  reach a target state from $s$ under the selected scheduler within the subsystem induced by the selected label set;
\item for each problematic state $s\in S$ a variable $r_s\in\Ireal$ for the encoding of increasing paths; and
\item for each problematic state-transition pair $(s,\edge)\in \probtrans_T$ and each successor
  state $s'\in\supp(\edge)$ a variable $t_{s,\edge,s'}\in\Iint$, where
  $t_{s,\edge,s'}=1$ implies that the values attached to the states
  increase along the branch $(s,\edge,s')$, \ie $r_s < r_{s'}$.
\end{itemize}

\noindent Let $w_{\min}=\min\bigl\{w(\ell)\,|\,\ell\in\Lab\wedge w(\ell)>0\bigr\}$ 
be the smallest positive weight that is assigned to any label. The
MILP for the smallest critical label set problem is then as follows:
\begin{subequations}
{\allowdisplaybreaks
\begin{align}
 \text{minimize}\quad & -\frac12 w_{\min}\cdot p_{\sinit} + \sum_{\ell\in\Lab} w(\ell)\cdot x_{\ell} 
      \label{eq:milp:obj} \\
 \text{such that}\quad & \notag \\
      & p_{\sinit}>\lambda 
      \label{eq:milp:bound} \\
 \forall s\in T\suchthat\quad 
      & p_s = 1 
      \label{eq:milp:target} \\
 \forall s\in S\setminus T\suchthat\quad 
      & \sum\limits_{\edge\in P(s)} \sigma_{s,\edge} \leq 1 
      \label{eq:milp:scheduler} \\
 \forall s\in S\setminus T\suchthat \quad 
      & p_s\leq \sum_{\edge\in P(s)}\sigma_{s,\edge} 
      \label{eq:milp:prob_sched} \\
 \forall s\in S\setminus T\suchthat \forall \edge\in P(s)\suchthat \forall s'\in\supp(\edge)\suchthat& \forall \ell\in\LT(s,\edge,s')\suchthat\notag \\ & p_{s,\edge,s'} \leq x_{\ell}
      \label{eq:milp:branch_label} \\
 \forall s\in S\setminus T\suchthat \forall \edge\in P(s)\suchthat \forall s'\in\supp(\edge)\suchthat\quad & p_{s,\edge,s'} \leq \edge(s')\cdot p_{s'}
      \label{eq:milp:branch_prob} \\
 \forall s\in S\setminus T\suchthat \forall \edge\in P(s)\suchthat \quad 
      & p_s\leq  (1-\sigma_{s,\edge}) + \sum_{s'\in\supp(\edge)} p_{s,\edge,s'} 
      \label{eq:milp:prob_comp} \\
\forall (s,\edge)\in\probtrans_T\suchthat\quad 
      & \sigma_{s,\edge} = \sum_{s'\in\supp(\edge)} t_{s,\edge,s'}  
      \label{eq:milp:reach1} \\
\forall (s,\edge)\in\probtrans_T\suchthat\forall s'\in\supp(\edge)\suchthat\quad 
      & r_s < r_{s'} + (1-t_{s,\edge,s'})
      \label{eq:milp:reach2}
\end{align}}
\end{subequations}

We first explain the constraints in
lines~\eqref{eq:milp:bound}--\eqref{eq:milp:reach2} of the MILP, which
describe a critical label set. First, we ensure that the
probability of the initial state exceeds the probability bound
$\lambda$~\eqref{eq:milp:bound}. The probability of target states is
set to 1~\eqref{eq:milp:target}. For
each state $s\in S\setminus T$ the scheduler selects at most one
transition, encoded by setting at most one scheduler variable 
$\sigma_{s,\edge}\in P(s)$ to $1$~\eqref{eq:milp:scheduler}.
Note that there may be states where no transition is chosen. In this
case the probability of a state is set to 0~\eqref{eq:milp:prob_sched}.
The next two constraints describe the probability contribution of an
edge $\edge\in P(s)$ from $s$ to $s'$: If a
label $\ell\in\LT(s,\edge,s')$ is not contained in the selected 
subset, the probability of the branch is set to 0~\eqref{eq:milp:branch_label}.
Otherwise this constraint is satisfied for all possible values of
$p_{s,\edge,s'}\in\Ireal$. Then the following constraint~\eqref{eq:milp:branch_prob} 
imposes an upper bound on the contribution of this branch, namely
the probability $\edge(s')$ of this branch times the probability of
the successor state $s'$. Constraint~\eqref{eq:milp:prob_comp}
is trivially satisfied if $\sigma_{s,\eta} = 0$, \ie if the scheduler 
does not select the current transition. Otherwise the probability
$p_s$ of state $s$ is at most the sum of the probabilities of its
outgoing branches.

The reachability of at least one deadlocking or non-problematic state is ensured by
\eqref{eq:milp:reach1} and \eqref{eq:milp:reach2}. First, if a
problematic transition $\edge$ of a state $s$ is selected by the scheduler then
exactly one transition branch flag must be activated. Second, for all paths along activated
branches of problematic transitions, an increasing order on the problematic states is enforced.
Because of this order, no problematic states can be revisited on an
increasing path which enforces the final reachability of a non-problematic or a deadlocking
state.

These constraints assure that each satisfying assignment of the 
label variables $x_{\ell}$ corresponds to a critical label set. By minimizing
the total weight of the selected labels we obtain a smallest critical label set.
By the additional term $-\frac{1}{2}w_{\min}\cdot p_{\sinit}$ we obtain not
only a smallest critical label set but one with maximal probability. The
coefficient $-\frac{1}{2}w_{\min}$ is needed to ensure that the benefit
from maximizing the probability is smaller than the loss by adding an
additional label. Please note that any coefficient $c$ with $0 < c < w_{\min}$
could be used.

\subsection{Size of the MILP}
The number of integer variables in this MILP is in $O(l+m)$,
the number of real variables in $O(n+m)$, and the number of
constraints in $O(n + l\cdot m)$ where $l = |\Lab|$ is the number of labels, 
$n = |S|$ the number of states, and $m$ the number of branches of PA $\PA$, \ie
$m = \bigl|\{ (s,\edge,s')\,|\,s\in S, \edge\in P(s), s'\in\supp(\edge)\}\bigr|$.

\begin{rem}
  \label{rem:special_case}
  In case the labeling $L(s,\edge,s')$ does not depend on the
  successor states $s'$, but only on the state $s$ and the selected
  transition $\edge$ or even only on $s$, then constraints
  \eqref{eq:milp:branch_label}--\eqref{eq:milp:prob_comp} can be 
  simplified. See~\cite{wimmer-et-al-qest-2013} for details.
\end{rem}

\subsection{Correctness of the MILP}
\label{subsec:proof}

For the correctness of the MILP formulation
\eqref{eq:milp:obj}--\eqref{eq:milp:reach2} we need to show that for
each critical label set there is a satisfying assignment of the MILP and
that from each satisfying assignment of the MILP one can construct a
critical label set. As setting we have again $\PA$, $T$, $L$, $\Lab$ and $\PAind$ for $\Lab'\subseteq \Lab$.

\begin{lem}
  \label{lemma:scl_to_ass}
  For each critical label set $\Lab'\subseteq\Lab$ there is an assignment
  $\nu$ of the MILP variables with $\nu(x_\ell)=1$ iff $\ell\in\Lab'$ such 
  that the constraints~\eqref{eq:milp:bound}--\eqref{eq:milp:reach2} are satisfied.
\end{lem}

\proof Let $\Lab'\subseteq\Lab$ be a critical label set. Then
$\probmax_{\PAind}(\sinit,\finally{T})>\lambda$ and a
deterministic memoryless scheduler $\sigma$ for $\PAind$ exists with
$\prob_{\PAind^\sigma}(\sinit,\finally{T}) =
\probmax_{\PAind}(\sinit,\finally{T})>\lambda$ and $s\in\dom(\sigma)$
iff $\probmax_{\PAind}(s,\finally{T})>0$, for all $s\in S\setminus T$.
Let $G=(V,E)$ be the digraph with $V=\dom(\sigma)\cup T$ and
$E=\{(s,s')\in V\times V\, | \, s'\in\supp(\sigma(s))\}$. 
Now consider a smallest (edge-minimal) subgraph $G'=(V,E')$ of $G$
containing for each state $s\in V$ a path from $s$ to $T$.  Due to
minimality, $G'$ is loop-free and contains for each state $s\in
V\setminus T$ exactly one outgoing edge.  We set
\[
\begin{array}{lll@{\qquad}lll}
\nu(x_\ell)&=&\left\{\begin{array}{l@{\quad}l} 1 & \text{if } \ell\in\Lab',\\ 0 & \text{otherwise;}\end{array} \right. &
\nu(\sigma_{s,\edge}) &=& \left\{ \begin{array}{l@{\quad}l} 1 & \text{if $s\in\dom(\sigma)$ and $\sigma(s) = \edge$,}\\ 0 & \text{otherwise;}  \end{array} \right. \\ \\
\nu(p_s) &=& \prob_{\PAind^\sigma}(s,\finally{T}); &
\nu(p_{s,\edge,s'}) &=& \left\{\begin{array}{l@{\quad}l} \edge(s')\cdot\nu(p_{s'}) & \text{if } \LT(s,\edge,s')\subseteq\Lab', \\ 
   0 & \text{otherwise;} \end{array}\right.
\end{array}
\]
\[
\begin{array}{lcl}
\nu(t_{s,\edge,s'})&=&\left\{\begin{array}{l@{\quad}l} 1& \text{if $s\in
  V\cap\probstates_T \text{ and } (s,s')\in E' \text{ and } \sigma(s)=\edge$,}\\
  0 & \text{otherwise;}\end{array}\right.\\ \\
  \nu(r_s) &=& \left\{\begin{array}{l@{\quad}l}\frac{1}{2}\nu(r_{s'}) & \text{if $s\in V\cap\probstates_T \text{ and} (s,s')\in E'$,}\\1 & \text{otherwise}.\end{array}\right.
\end{array}
\]
We now systematically check the constraints \eqref{eq:milp:bound} through \eqref{eq:milp:reach2}:
\begin{enumerate}[label=(3.1\alph*),start=2]
 \item
  is satisfied by $\nu$ because
  $\nu(p_{\sinit}) = \prob_{\PAind^\sigma}(\sinit,\finally{T})>\lambda$.
 \item
  holds because
  $\prob_{\PAind^\sigma}(s,\finally{T})=1$ for all target states $s\in T$. 
 \item
   holds since the deterministic memoryless
  scheduler $\sigma$ selects at most one transition in each state.
\item
  is trivially satisfied if
  $\nu(\sigma_{s,\edge})=1$ for some $\edge\in P(s)$. Otherwise, if no
  action is chosen, $s$ is a deadlock state and the probability
  $\nu(p_s)$ to reach a target state is $0$.
 \item
  is satisfied as
  $\nu(p_{s,\edge,s'})$ is defined to be $0$ if 
  $\ell\in\LT(s,\edge,s')$ for some $\ell\not\in\Lab'$. 
 \item
  holds by the definition of
  $\nu(p_{s,\edge,s'})$.
 \item
  is trivially satisfied if 
  $\nu(\sigma_{s,\edge}) = 0$. In case $\nu(\sigma_{s,\edge}) = 1$, 
  the constraint reduces to
  $p_s \leq\sum_{s'\in\supp(\edge)} p_{s,\edge,s'} \leq \sum_{s'\in\supp(\edge)} \edge(s')\cdot p_{s'}$ with $\edge=\sigma(s)$.
  It is satisfied if $\nu(p_s) = 0$.
  Otherwise, since $\nu(p_s)$ is the reachability probability of $T$ in 
  $\PAind^\sigma$, it satisfies the following 
  equation~\cite[Theorem~10.19]{BK08}:
  \[
    \nu(p_s) = \sum_{s'\in\supp(\edge)} \edge(s')\cdot\nu(p_{s'})
             = \sum_{s'\in\supp(\edge)} \nu(p_{s,\edge,s'}).
  \]
  $\PAind$ contains exactly those branches $(s,\edge,s')$ of 
  $\PA^{\sigma}$ for which $\LT(s,\edge,s')\subseteq\Lab'$ and
  therefore $\nu(p_{s,\edge,s'}) = \edge(s')\cdot\nu(p_{s'})$. For all 
  other branches $(s,\edge,s')$ in $\PA^{\sigma}$, but not in 
  $\PAind^{\sigma}$, $\nu(p_{s,\edge,s'})=0$ holds. Hence we have
  $
  \nu(p_s) = \sum_{s'\in\supp(\edge)} \nu(p_{s,\edge,s'})
  $
  and \eqref{eq:milp:prob_comp} is satisfied.
\item
  holds if
  $\nu(\sigma_{s,\edge})=0$, since in this case by definition either
  $s\notin\dom(\sigma)$ or $\edge\not=\sigma(s)$ and therefore
  $\nu(t_{s,\edge,s'})=0$ for all $s'\in S$.  Otherwise
    $\nu(\sigma_{s,\edge})=1$, \ie $\sigma(s)=\edge$. By the
    construction of $G'$ there is exactly one $s'\in\supp(\edge)$ with
    $\nu(t_{s,\edge,s'})=1$.
\item
  is straightforward if
  $\nu(t_{s,\edge,s'})=0$. Otherwise by definition
  $r_s=\frac{1}{2}r_{s'}$, and since $\nu(r_s),\nu(r_{s'})>0$, the inequality
  holds.\qed
\end{enumerate}

\begin{lem}
  \label{lemma:ass_to_scl}
  Let $\nu$ be a satisfying assignment of the MILP 
  \eqref{eq:milp:bound}--\eqref{eq:milp:reach2}. Then
  $\Lab' = \{ \ell\in\Lab\,|\,\nu(x_\ell)=1\}$ is a critical label set.
\end{lem}

\proof
  \def\nuopt{\nu_{\mathrm{opt}}}
  \def\ximax{\xi^{\mathrm{max}}}

  Let $\nu$ be a satisfying assignment and 
  $\Lab' = \{\ell\in\Lab\,|\,\nu(x_\ell)=1\}$ the induced label set.
  We define the scheduler
  $\sigma:S\partialto\Act\times\SubDistr(S)$ by $\dom(\sigma)=\{s\in
  S\, | \, \exists \edge\in P(s)\suchthat \nu(\sigma_{s,\edge})=1
  \wedge \exists s'\in\supp(\edge).\ L(s,\edge,s')\subseteq \Lab'\}$
  and for each $s\in\dom(\sigma)$ we set $\sigma(s)=\edge$ with
  $\nu(\sigma_{s,\edge})=1$. Due
  to constraint~\eqref{eq:milp:scheduler} there is at most one
  transition $\edge\in P(s)$ for $\sigma_{s,\edge}=1$. Therefore the
  scheduler is well defined. If $s\notin\dom(\sigma)$ then
  $s$ is a deadlock state under $\sigma$ with no outgoing transition. 

  Let $\sigma_{|\Lab'}:S\partialto\Act\times\SubDistr(S)$ result
  from $\sigma$ by removing branches whose labels are not included in
  $\Lab'$, \ie $\dom(\sigma_{|\Lab'})=\dom(\sigma)$ and for each $s\in\dom(\sigma_{|\Lab'})$ and $s'\in S$,
  $\sigma_{|\Lab'}(s)(s')=\sigma(s)(s')$ if $L(s,\sigma(s),s')\subseteq \Lab'$, and
  $\sigma_{|\Lab'}(s)(s')=0$ otherwise.

  Let $U$ be the set of states from which $T$ is not reachable in
  $\dtmc$\footnote{Note that the order of operations is not arbitrary here. We have $(\PA^\sigma)_{|\Lab'}=(\PAind)^{\sigma_{|\Lab'}}$.}, $D$ the deadlock states in $U$, and $R$ the states in $U$
  whose scheduled transitions got reduced by removing some branches
  due to the selected label set:
  \begin{align*}
    U &= \{s\in S\,|\,\text{$T$ is unreachable from $s$ in $\dtmc$}\}\\
    D &= U\setminus\dom(\sigma_{|\Lab'})\\
    R &= \{s\in U\cap\dom(\sigma_{|\Lab'})\, | \, \sigma(s)\not=\sigma_{|\Lab'}(s) \}\, .
  \end{align*}

The reachability probabilities $q_s=\prob_{\dtmc}(s,\finally{T})$ are the
unique solution of the following linear equation
system~\cite[Theorem~10.19]{BK08}:
  \begin{gather}
    \label{eq:mc_dtmc}
    q_s = \begin{cases}
        1 & \text{for $s\in T$,}\\
        0 & \text{for $s\in U$,} \\
        \sum_{s'\in S}\sigma_{|\Lab'}(s)(s')\cdot q_{s'} & \text{otherwise.}
        \end{cases}
  \end{gather}
  This equation system is well defined, since, if $\sigma_{|\Lab'}(s)$ is
  undefined, either $s$ is a target state or the target states are
  unreachable from $ s$.
  In the following we prove
  \begin{align}
    \nu(p_s) &= 1 && \text{for $s\in T$,}       \label{eq:ps_target}\\
    \nu(p_s) &= 0 && \text{for $s\in U$,}       \label{eq:ps_unreach}\\
    \nu(p_s) & \leq q_s && \text{otherwise}\, . \label{eq:ps_else}
  \end{align}
  Thus $\nu(p_s)\leq
  q_s$ for each $s\in S$. With \eqref{eq:milp:bound} we get
  $q_{\sinit}>\lambda$, \ie $\Lab'$ is critical.

\medskip

\noindent It remains to show that \eqref{eq:ps_target}--\eqref{eq:ps_else} hold.

\begin{enumerate}[label=(3.\arabic*),start=3]

\item
  is straightforward for target states $s\in T$ due to \eqref{eq:milp:target}.

\item
  First we observe that from all states
  $s\in U$ a state in $D\cup R$ is reachable: Equations
  \eqref{eq:milp:reach1} and \eqref{eq:milp:reach2} assure that from
  each problematic state in $U$ we can reach a state from $D\cup R$
  (proof by induction over the problematic states $s\in\probstates_T$
  with decreasing values $r_s$).  From the non-problematic states in
  $U$ the target states $T$ are reachable in $\PA$ under each
  scheduler. Therefore, the unreachability of $T$ from those states in
  $\dtmc$ is due to the selected label set, where certain branches on each
  path leading to $T$ are not available any more.  Thus also from each non-problematic state in
  $U$ we can reach a state in $D\cup R$.

Now we show that $\nu(p_s)=0$ for all $s\in U$. Assume the opposite
and let $s\in U$ with $\nu(p_s)=\ximax=\max\{\nu(p_{s'})\, | \, s'\in
U\}>0$. Then $s\in\dom(\sigma_{|\Lab'})$ by Equations
\eqref{eq:milp:prob_sched}--\eqref{eq:milp:prob_comp}, and for $\sigma_{|\Lab'}(s)=\edge$ we get:
  \begin{gather}
    \label{eq:ximax}
    \begin{split}
     \ximax\  =\ \nu(p_s)\ &\leq  \quad \sum_{s'\in\supp(\edge)} \edge(s')\cdot \nu(p_{s'}) 
            \quad \leq \quad \sum_{s'\in\supp(\edge)} \edge(s')\cdot \ximax \\
            & = \quad \ximax\cdot\sum_{s'\in\supp(\edge)}\edge(s') 
            \quad \leq \quad \ximax\ .
     \end{split}
  \end{gather}
  Therefore all inequalities have to hold with equality. Since
  $\ximax$ is assumed to be positive, this is possible only if
  $\sum_{s'\in\supp(\edge)}\edge(s')=1$, \ie $s\in U\setminus R$, and
  $\nu(p_{s'}) = \ximax$ for all $s'\in\supp(\edge)$.  By induction we
  conclude that $\nu(p_{s'}) = \ximax$ and $s'\in U\setminus R$ for
  all states $s'$ that are reachable from $s$ under $\sigma_{|\Lab'}$.  We
  know that from each $s\in U$ either a state $s'\in D$ or a state
  $s'\in R$ is reachable. For the former case $s'\in D$, from
  \eqref{eq:milp:prob_sched}--\eqref{eq:milp:prob_comp} we imply
  $\nu(p_{s'})=0$, contradicting to $\nu(p_{s'}) = \ximax > 0$. In the
  latter case $s'\in R$, the definition of $R$ implies
  $\sum_{s''\in\supp(\sigma_{|\Lab'}(s'))}\sigma_{|\Lab'}(s')(s'') < 1$, contradicting to
  $\sum_{s''\in\supp(\sigma_{|\Lab'}(s'))}\sigma_{|\Lab'}(s')(s'')=1$. Therefore our
  assumption was wrong and we have proven $\nu(p_s)=0$ for each $s\in
  U$.

\item
  Finally we show that 
  $\nu(p_s)\leq q_s$ for each $s\in S\setminus (T\cup U)$.  The
  constraints \eqref{eq:milp:branch_label}--\eqref{eq:milp:prob_comp} can be simplified for the
  chosen action $\sigma_{|\Lab'}(s)=\edge$ to:
  \begin{gather}
    p_s \leq\sum_{s'\in\supp(\edge)}\edge(s')\cdot p_{s'}
    \label{eq:reach}
  \end{gather}
  Let $\nuopt$ be a satisfying assignment such that $\nuopt(p_{s})$ is
  maximal among all satisfying assignments (this maximum exists, since
  the set of satisfying assignments is compact).
  We claim that for all $s'\in S\setminus (T\cup U)$ reachable from $s$ in $\dtmc$, 
  Equation \eqref{eq:reach} is satisfied by $\nuopt$ 
  with equality. Assume the converse is true, \ie there is a state 
  $s'\in S\setminus (T\cup U)$ that is reachable from $s$ in $\dtmc$ such that $\sigma_{|\Lab'}(s)=\edge$ and 
  \[
    0 < \varepsilon =
      \biggl(\sum\limits_{s''\in\supp(\edge)} \edge(s'')\cdot \nuopt(p_{s''})\biggr) 
      - \nuopt(p_{s'})\ .
   \]
   Let $s = s_0 \edge_0 s_1 \edge_1\ldots s_n = s'$ be an
   acyclic path in $\dtmc$ from $s$ to $s'$. We could increase the
   value $\nuopt(p_{s_n})$ by at least $\varepsilon_n=\varepsilon$
   (more, if $p_{s_n}$ also appears on the right-hand side; note that
   $0 \leq \edge_{i}(s_i) < 1$ holds for all $i=0,\ldots,n$).  This would
   not violate any inequality, since in the inequalities for the other
   states $p_{s_n}$ appears only in upper bounds on the right-hand sides with a non-negative coefficient.
   Assume that, for some $i\leq n$, we have increased the value of
   $s_i$ by $\varepsilon_i$. Then the right-hand side of the
   inequality for $s_{i-1}$ increases by at least
   $\edge_{i-1}(s_i)\cdot\varepsilon_i>0$. Therefore we could also
   increase the value of $p_{s_{i-1}}$ by
   $\edge_{i-1}(s_i)\cdot\varepsilon_i$.  This could be continued along
   the path back to $s=s_0$, whose value could be increased by
   $\varepsilon_0=\varepsilon\cdot\prod_{i=0}^{n-1}\edge_i(s_{i+1})>0$.
   But then $\nuopt(p_{s})$ would not be optimal, contradicting our
   assumption $\varepsilon>0$. 

   This means, the inequalities for all states that are
   reachable from $s$ are satisfied with equality for $\nuopt$, in other words,
   $\nuopt$ encodes the solution $\nuopt(p_s)=q_s$ to \eqref{eq:mc_dtmc}.
   Since $\nuopt$ is maximal
   for $s$, all other assignments satisfy $\nu(p_s)\leq q_s$.\qed

\end{enumerate}

\begin{thm}
  \label{th:correctness}
  The MILP given in \eqref{eq:milp:obj}--\eqref{eq:milp:reach2} yields 
  a smallest critical label set.
\end{thm}

\proof According to Lemmas~\ref{lemma:scl_to_ass} and
\ref{lemma:ass_to_scl}, for each critical label set $\Lab'$ there is a
satisfying assignment $\nu$ and vice versa. With
$\Lab'=\{\ell\in\Lab\, |\, \nu(x_{\ell})=1\}$ and
$w(\Lab')=\sum_{\ell\in\Lab'}w(\ell)=\sum_{\ell\in\Lab} w(\ell)\cdot
\nu(x_{\ell})$, for the objective function
  \[
     w(\Lab')-w_{\min}<-\frac12 w_{\min}\cdot \nu(p_{\sinit}) + w(\Lab') < w(\Lab')
  \]
  holds.
  By minimizing the objective function, we obtain a smallest critical label set.
\qed


\subsection{Optimizations}
\label{subsec:optimizations}

The constraints of the MILP describe \emph{critical label sets},
whereas \emph{minimality} is enforced by the objective function. In
this section we describe how some additional constraints can be
imposed, which explicitly exclude variable assignments that are
either not optimal or encode label sets that are also encoded by other
assignments. The branch\,\&\,bound methods used for solving MILPs~\cite{Sch86}
obtain lower bounds on the optimal solution by solving a linear relaxation
of the problem. The lower bounds are used to prune branches of the
search space which cannot contain an optimal solution. Adding constraints 
that are redundant regarding the feasible solutions can improve the relaxation, 
yield larger lower bounds and let the solver thus prune more branches of the
search space. This can reduce the computation time significantly in spite
of the larger number of constraints.

\paragraph{\bfseries Scheduler cuts}
We want to exclude solutions of the constraint set for which a
non-deadlocking state $s$ has only \emph{deadlocking successors} under
the selected scheduler. Note that such solutions would define $p_s=0$.
We add for all $s\in S\setminus T$ and all $\edge\in P(s)$ with
$\supp(\edge)\cap T=\emptyset$ the constraint
  \begin{gather}
    \label{eq:opt:sched_fwd}
    \sigma_{s,\edge}\ \leq\ \sum_{s'\in\supp(\edge)\setminus \{s\}}\ \sum_{\edge'\in P(s')}\sigma_{s',\edge'}\ .
  \end{gather}
Analogously, we require for each
non-initial state $s$ with a selected action-distribution pair
$\edge\in P(s)$ that there is a selected action-distribution pair leading
to $s$. Thus, we add for all states $s\in S\setminus\{\sinit\}$ the
constraint
  \begin{gather}
    \label{eq:opt:sched_bwd}
    \sum_{\edge\in P(s)\ }\sigma_{s,\edge}\leq 
    \sum_{s'\in \{s''\in S\,|\, s''\not=s\wedge
      \exists \edge\in P(s'')\suchthat s\in\supp(\edge)\}}
      \quad \sum_{\edge'\in \{\edge''\in P(s')\,|\,
      s\in\supp(\edge'')\}}\sigma_{s',\edge'}\ .
 \end{gather}
As special cases of these cuts, we can encode that the initial state
has at least one activated outgoing transition and that at least one
of the target states has a selected incoming transition. These
special cuts come with very few additional constraints and often have
a great impact on the solution times.

\paragraph{\bfseries Label cuts}
In order to guide the solver to select the correct combinations of
\emph{labels} and \emph{scheduler} variables, we want to enforce that for every
selected label $\ell$ there is at least one scheduler variable
$\sigma_{s,\edge}$ activated such that $\ell\in \bigcup_{s'\in\supp(\edge)}\LT(s,\edge,s')$:
\begin{gather}
  \label{eq:opt:lc}
  x_\ell\ \leq\ \sum_{s\in S}\quad\sum_{\edge\in\{\edge'\in P(s)\,|\,\exists s'\in\supp(\edge')\suchthat\ell\in \LT(s,\edge',s')\}}\sigma_{s,\edge}\ .
\end{gather}

\paragraph{\bfseries Synchronization cuts}

While scheduler and label cuts are applicable to the general smallest
critical label set problem, synchronization cuts take the proper
\emph{synchronization of commands} into account and are
applicable for the computation of smallest critical command sets only.

Let $M_i, M_j$ ($i\neq j$) be two modules which synchronize on action $\act$,
$c$ a command of $M_i$ with action $\act$, and $C_{j,\act}$ the set of commands 
with action $\act$ in module $M_j$. The following constraint ensures that
if command $c$ is selected by activating the variable $x_{l_c}$ then at least
one command $d\in C_{j,\act}$ is selected, too. 
\begin{gather}
  \label{eq:opt:sync}
  x_{\ell_c} \ \leq \ \sum_{d\in C_{j,\act}} x_{\ell_d} \ .
\end{gather}
\noindent Similar constraints can be
formulated for minimization of command branches.


\subsection{Further Simplification of Counterexamples}
\label{sec:simplification}

Given a PA $\PA$ as a PRISM program and a violated reachability property
$\pctlProb{\leq\lambda}{\finally{T}}$, we propose to extract a counterexample
as follows:
\begin{enumerate}
\item remove all \emph{commands} which are not necessary for a violation of the property,
\item remove all unnecessary \emph{branches} of the remaining commands, and
\item reduce the \emph{domains} of the variables as much as possible.
\end{enumerate}
These simplification steps are special cases of the SCL problem, as
described in the previous section. As an alternative to the last step, 
the \emph{intervals} of the variables can be reduced, which is also an application 
of the SCL problem, but with additional constraints.

\subsection{Reduction of Variable Intervals}
\label{subsec:interval_reduction}

Using SCL to reduce variable domains as described in Section~ \ref{subsec:scl}
can yield domains that are not intervals anymore. If intervals are desired,
additional constraints are necessary to keep the domains connected.

For each variable $\var\in\Vars$ we encode an interval
$[l_\var,u_\var]\subseteq\dom(\var)\subseteq\mathbb{Z}$ by introducing two
additional variables $h_{\var,v}^u,h_{\var,v}^l\in\Iint$ for each
$v\in\dom(\var)$. The intuition is that $h_{\var,v}^u = 1$ iff $v>u_\var$,
and $h_{\var,v}^l=1$ iff $v<l_\var$. The remaining values $v\in\dom(\var)$,
for which $h^u_{\var,v}=0$ and $h^l_{\var,v}=0$ hold, form the
interval. We add the following additional constraints to the MILP
\eqref{eq:milp:obj}--\eqref{eq:milp:reach2}:
\begin{subequations}
\begin{align}
  \forall \var\in\Vars\suchthat \forall v\in\dom(\var)\suchthat v\neq\min\bigl(\dom(\var)\bigr)\suchthat \quad & 
     h^l_{\var,v} \leq h^l_{\var,v-1}
     \label{eq:intred:lb} \\
  \forall \var\in\Vars\suchthat \forall v\in\dom(\var)\suchthat v\neq\max\bigl(\dom(\var)\bigr)\suchthat \quad & 
     h^u_{\var,v} \leq h^u_{\var,v+1}
     \label{eq:intred:ub} \\
  \forall \var\in\Vars\suchthat \forall v\in\dom(\var)\suchthat \quad &
     h^l_{\var,v} + h^u_{\var,v} + x_{\ell_{\var,v}} = 1
     \label{eq:intred:lab}
\end{align}
\end{subequations}
The first constraint takes care that, if a value $v$ is neglected (\ie $h^l_{\var,v}=1$), also 
the value $v-1$ is neglected. The same holds for $h^{u}_{\var,v}$ and the successor value $v+1$
in \eqref{eq:intred:ub}. Constraint \eqref{eq:intred:lab} connects the decision variables $x_{\ell_{\var,v}}$
for the labeling with the auxiliary variables $h^l_{\var,v}$ and $h^u_{\var,v}$: Exactly
one of these three variables has to be set to 1---either $v$ is below the lower bound ($h^l_{\var,v}=1$) or
above the upper bound ($h^u_{\var,v}=1$), or the label $\ell_{\var,v}$ is contained in the
computed label set.

\section{Experiments}
\label{sec:experiments}

We have implemented the described techniques in \cpp{} using the MILP solver
\tool{Gurobi}~\cite{gurobi}. The experiments were performed on an Intel\TReg{}
Xeon\TReg{} CPU E5-2643 with 3.3~GHz clock frequency and 32~GB of main memory,
running Ubuntu 12.04 Linux in 64~bit mode. We focus on the minimization of the
number of commands needed to obtain a counterexample and simplify them by
deleting a maximum number of branchings and variable values. We do not consider symmetries in the
models. We ran our tool with two threads in parallel and aborted any experiment
which did not finish within 10~min (1200~CPU seconds). We conducted a number of
experiments with benchmarks that are publicly available on the web page of
\tool{\tool{PRISM}}~\cite{KNP12b}. We give a brief overview of the used models.

\noindent\mbox{\hfill $\blacktriangleright$ \texttt{coin}-$N$-$K$}~ \cite{KNS01a}
models the shared coin protocol as in Example~\ref{ex:coin}. The
protocol is parameterized by the number $N$ of involved processes, which collectively undertake a random walk by flipping an unbiased coin and, depending on the outcome, incrementing or decrementing a shared counter. If the counter reaches a value greater than $KN$ for an
integer constant $K > 1$ then the decision is heads, if it is less than $-KN$ then tails. We consider the property
$\mathcal{P}_{\leq\lambda}\bigl(\finally{(\mathrm{finished}\land\mathrm{all\_coins\_equal\_0})}\bigr)$,
which is satisfied if the probability to finish the protocol with all
coins equal to $0$ is at most $\lambda$. 

For $N=K=2$, this probability is $0.5556$. To show the
applicability of our approach we introduce a ``bug'' by having a
biased coin where the probability for $\texttt{coin=0}$ is $0.8$ for
all processes. The probability is now $0.9999$. If
we search for a smallest critical command set for $\mathcal{P}$ with a
probability bound of $0.5$, which is the expected scenario, the
command~\ref{coin_increment} as in Example~\ref{ex:coin} is not
chosen. That means to observe faulty behavior it is not necessary to
ever increment the counter. This gives us the hint that the fault
is caused by the flipping command~\ref{coin_flip}.

\noindent\mbox{\hfill $\blacktriangleright$ \texttt{wlan}-$B$-$C$}
models the two-way handshake mechanism of the IEEE 802.11 Wireless LAN
protocol. Two stations try to send data, but run into a
collision. Therefore they enter the randomized exponential backoff
scheme. The parameter $B$ denotes the maximal allowed value of the
backoff counter. We check the property
$\mathcal{P}_{\leq\lambda}\bigl(\finally{(\mathrm{num\_collisions} =
  C)}\bigr)$ putting an upper bound on the probability that a maximal
allowed number $C$ of collisions occur.

\noindent\mbox{\hfill $\blacktriangleright$ \texttt{csma}-$N$-$C$}
concerns the IEEE 802.3 CSMA/CD network protocol. $N$ is the number of
processes that want to access a common channel, $C$ is the maximal
value of the backoff counter. We check 
$\mathcal{P}_{\leq\lambda}(\neg\mathrm{collision\_max\_backoff}\,U\,\mathrm{delivered})$ 
expressing that the probability that all stations successfully 
send their messages before a collision with maximal backoff occurs 
is at most $\lambda$.

\noindent\mbox{\hfill $\blacktriangleright$ \texttt{fw}-$N$}
models the Tree Identify Protocol of the IEEE 1394 High Performance
Serial Bus (called ``FireWire'') \cite{Stoe03}.  It is a leader
election protocol which is executed each time a node enters or leaves
the network. The parameter $N$ denotes the delay of the wire as
multiples of 10~ns. We check 
$\mathcal{P}_{\leq\lambda}(\finally{\mathrm{leader\_elected}})$, \ie that
the probability of finally electing a leader is at most $\lambda$.

Some statistics of the models for different parameter values are shown
in Table~\ref{tab:model_statistics}. The columns contain the name of
the model (``Name''), its number of states (``\#states''), 
transitions (``\#trans''), modules (``\#mod''), commands (``\#comm''), 
branches of the commands (``\#upd''), variables (``\#var''), 
and different variable values (``\#val''), \ie $\sum_{\var\in\Vars}|\dom(\var)|$.
The values in braces are the number of commands and command branches,
respectively, that are relevant for the considered property. 
Column ``$\probmax(\sinit,\finally{T})$'' contains the
reachability probability and column ``$\lambda$'' the probability bound. The last
column ``MCS''shows the number of states in the minimal critical subsystem,
\ie the smallest subsystem of the PA such that the probability to
reach a target state inside the subsystem is still above the
bound. Entries which are marked with a star correspond to the
smallest critical subsystem we could find within the time bound of
10~min using our tool \tool{LTLSubsys}~\cite{wimmer-et-al-tacas-2012}, 
but they are not necessarily optimal.

\begin{table}[tb]
\centering
\caption{Model statistics}
\label{tab:model_statistics}
\scalebox{0.9}{%
  \setlength{\tabcolsep}{1.5mm}
  \setlength{\tabcolsep}{3pt}
\begin{tabular}{l@{\qquad}r@{\quad}rcrrrrn{1}{5}n{1}{3}r}
\toprule
Model   & \#states & \#trans & \#mod & \#comm & \multicolumn{1}{c}{\#upd} & \#var & \#val 
        & \multicolumn{1}{c}{$\probmax(\sinit,\finally{T})$} & \multicolumn{1}{c}{$\lambda$} & MCS \\
\midrule
coin2-1 & 144  &  252  & 2 & 14 (12) & 16 (12) & 5 & 19 & 0.6  & 0.4      & 13\phantom{${}^\ast$} \\
coin2-2 & 272  &  492  & 2 & 14 (12) & 16 (12) & 5 & 23 & 0.5556  & 0.4   & 25${}^\ast$ \\
coin2-4 & 528  &  972  & 2 & 14 (12) & 16 (12) & 5 & 31 & 0.52940  & 0.4  & 55${}^\ast$ \\
coin2-5 & 656  &  1212 & 2 & 14 (12) & 16 (12) & 5 & 35 & 0.52379  & 0.4  & 67${}^\ast$ \\
coin2-6 & 784  &  1452 & 2 & 14 (12) & 16 (12) & 5 & 39 & 0.51998  & 0.4  & 83${}^\ast$ \\
csma2-2 & 1038 &  1282 & 3 & 34 (34) & 42 (42) & 11 & 94 & 0.875    & 0.5  & 540\phantom{${}^\ast$} \\
csma2-4 & 7958 &  10594& 3 & 38 (38) & 90 (90) & 11 & 122 & 0.99902  & 0.5  & 1769${}^\ast$ \\[.7ex]
fw01     & 1743 & 2197  & 4 & 68 (64) & 72 (68) & 10 & 382 & 1.0 & 0.5     & 412\phantom{${}^\ast$} \\
fw04     & 5452 & 7724  & 4 & 68 (64) & 72 (68) & 10 & 394 & 1.0 & 0.5     & 412${}^\ast$ \\
fw10    &17190 &29364  & 4 & 68 (64) & 72 (68) & 10 & 418 & 1.0 & 0.5     & 412${}^\ast$ \\
fw15    &33425 &63379  & 4 & 68 (64) & 72 (68) & 10 & 438 & 1.0 & 0.5     & 412${}^\ast$ \\[.7ex]
wlan0-2 & 6063 & 10619 & 3 & 70 (42) & 100 (72) & 13 & 91 & 0.18359  & 0.1  & 121\phantom{${}^\ast$}\\
wlan0-5 &14883 & 26138 & 3 & 70 (42) & 100 (72) & 13 & 94 & 0.00114  & 0.001 & 952${}^\ast$ \\
wlan2-1 &28597 & 57331 & 3 & 76 (14) & 114 (14) & 13 & 100 & 1.0      & 0.5  & 7\phantom{${}^\ast$} \\
wlan2-2 &28598 & 57332 & 3 & 76 (42) & 114 (72) & 13 & 101 & 0.18260  & 0.1  & 121${}^\ast$ \\
wlan2-3 &35197 & 70216 & 3 & 76 (42) & 114 (78) & 13 & 102 & 0.01793  & 0.01 & 514${}^\ast$ \\
wlan3-1 &96419 & 204743& 3 & 78 (14) & 130 (14) & 13 & 110 & 1.0      & 0.5  & 7\phantom{${}^\ast$} \\
wlan3-2 &96420 & 204744& 3 & 78 (42) & 130 (72) & 13 & 111 & 0.18359  & 0.1  & 121${}^\ast$ \\
\bottomrule
\end{tabular}
}
\end{table}

The results of our experiments computing a smallest critical command
set are displayed in Table~\ref{tab:command}. The first column ``Model''
contains the name of the model. The following two blocks contain the
results of runs without any cuts (cf.\ Section~\ref{subsec:optimizations}) 
and with the best combination of cuts: If there were cut combinations
with which the MILP could be solved within the time limit, we report
the fastest one. If all combinations timed out, we report the one that
yielded the largest lower bound.

For the block without cuts, we give the number of variables (``Var.'')
and constraints (``Constr.'') of the MILP, the computation time in seconds 
(``Time''), the memory consumption in MB (``Mem.''), the number of commands in
the critical command set (``$n$''), and, in case the time limit was
exceeded, a lower bound on the size of the smallest critical command
set (``lb''), which the solver obtains by solving a linear programming
relaxation of the MILP. An entry ``??'' for the number of commands means that
the solver was not able to find a non-trivial critical command set
within the time limit. For the best cut combination, the last four
columns specify the combination of cuts leading to the best running time.
Here the column ``$\sigma_f$'' corresponds to scheduler forward cuts
\eqref{eq:opt:sched_fwd}, ``$\sigma_b$'' to scheduler backward cuts
\eqref{eq:opt:sched_bwd}, ``$\ell$'' to label cuts \eqref{eq:opt:lc}, and
``$\|$'' to synchronization cuts \eqref{eq:opt:sync}. An entry ``$\surd$''
indicates that the corresponding constraints have been added to the MILP,
``$\times$'' that they were not used.

Although we ran into timeouts for many instances, in particular
without any cuts, in almost all cases a solution could be found within
the time limit. We suppose that also the solutions of the aborted
instances are optimal or close to optimal. It seems that the MILP
solver is able to quickly find good (or even optimal) solutions due to
sophisticated heuristics, but proving their optimality is hard. A
solution is proven optimal as soon as the objective value of the best
solution and the lower bound coincide. The additional cuts
strengthen this lower bound considerably. Further experiments have
shown that the scheduler forward cuts of~Eq.\eqref{eq:opt:sched_fwd} have the
strongest effect on the lower bound. Choosing good cuts consequently
enables the solver to obtain optimal solutions for more benchmarks.

\begin{table}[tb]
  \caption{Results for command minimization (TO = 600 seconds)}
  \label{tab:command}
  \centering
  \scalebox{0.85}{\addtolength{\tabcolsep}{0.1pt}\begin{tabular}{lrrrrrr@{\quad}rrrrcccc}
  \toprule
        & \multicolumn{6}{c}{no cuts} & \multicolumn{8}{c}{best cut combination} \\
  \cmidrule(lr){2-7}\cmidrule(lr){8-15}
  Model & Var. & Constr. & Time & Mem. & $n$ & lb & Time & Mem. & $n$ & lb & $\sigma_f$ & $\sigma_b$ & $\ell$ & $\|$ \\
  \midrule
  coin2-1 & 277 & 492 & \TO & 855 & 9 & 8 & 69.05 & 42 & 9 & \opt & $\surd$ & $\surd$ & $\surd$ & $\times$ \\
  coin2-2 & 533 & 1004 & \TO & 1140 & 9 & 6 & \TO & 1111 & 9 & 8 & $\times$ & $\times$ & $\surd$ & $\surd$ \\
  coin2-4 & 1045 & 2028 & \TO & 553 & 9 & 6 & \TO & 974 & 9 & 7 & $\times$ & $\surd$ & $\surd$ & $\surd$ \\
  coin2-5 & 1301 & 2540 & \TO & 768 & 9 & 5 & \TO & 554 & 9 & 6 & $\times$ & $\surd$ & $\surd$ & $\surd$ \\
  coin2-6 & 1557 & 3052 & \TO & 698 & 9 & 5 & \TO & 688 & 9 & 6 & $\times$ & $\surd$ & $\surd$ & $\times$ \\
  csma2-2 & 2123 & 5990 & 7.30 & 26 & 32 & \opt & 4.19 & 28 & 32 & \opt & $\times$ & $\times$ & $\times$ & $\surd$ \\
  csma2-4 & 15977 & 46882 & 196.11 & 215 & 36 & \opt & 77.43 & 261 & 36 & \opt & $\times$ & $\surd$ & $\surd$ & $\surd$ \\
  fw01 & 3974 & 13121 & \TO & 148 & 28 & 27 & 29.43 & 122 & 28 & \opt & $\surd$ & $\times$ & $\times$ & $\times$ \\
  fw04 & 13144 & 43836 & \TO & 604 & 28 & 22 & 103.37 & 296 & 28 & \opt & $\surd$ & $\surd$ & $\times$ & $\surd$ \\
  fw10 & 46282 & 153764 & \TO & 954 & 28 & 15 & 390.82 & 1102 & 28 & \opt & $\surd$ & $\times$ & $\times$ & $\surd$ \\
  fw15 & 96222 & 318579 & \TO & 1494 & 28 & 10 & \TO & 1861 & 28 & 19 & $\surd$ & $\surd$ & $\times$ & $\times$ \\
  wlan0-2 & 7072 & 6602 & \TO & 474 & 33 & 15 & \TO & 373 & 33 & 31 & $\surd$ & $\surd$ & $\surd$ & $\surd$ \\
  wlan0-5 & 19012 & 25808 & \TO & 1083 & \novalue & \novalue & \TO & 1083 & \novalue & \novalue & $\times$ & $\times$ & $\times$ & $\times$ \\
  wlan2-1 & 28538 & 192 & 1.12 & 45 & 8 & \opt & 0.82 & 45 & 8 & \opt & $\surd$ & $\surd$ & $\times$ & $\times$ \\
  wlan2-2 & 29607 & 6602 & \TO & 517 & 33 & 15 & \TO & 416 & 33 & 31 & $\surd$ & $\surd$ & $\times$ & $\surd$ \\
  wlan2-3 & 36351 & 18922 & \TO & 809 & 38 & 14 & \TO & 394 & 38 & 32 & $\surd$ & $\surd$ & $\times$ & $\surd$ \\
  wlan3-1 & 96360 & 192 & 1.98 & 142 & 8 & \opt & 1.98 & 142 & 8 & \opt & $\times$ & $\times$ & $\times$ & $\times$ \\
  wlan3-2 & 97429 & 6602 & \TO & 552 & 33 & 15 & \TO & 518 & 33 & 31 & $\surd$ & $\surd$ & $\times$ & $\surd$ \\
  \bottomrule
\end{tabular}
}
\end{table}

Table~\ref{tab:upvar} contains the results of the subsequent
simplification steps: To the \tool{PRISM} model corresponding to the smallest
critical command set we applied minimization of the commands' branches
and finally to its result the minimization of variable domains. For
both we give, as before, the computation time in seconds, the memory
consumption in megabytes, the resulting number of branches and
variable values, respectively, and, if a time out occured, the
computed lower bound. For both simplification steps we report only the
running times obtained using the best combination of cuts. In all
experiments that terminated within the time limit, the branch and
variable domain minimization were faster than the previous command
selection using the same combination of cuts.

We have also ran the two simplification steps directly on the original
\tool{PRISM} model with all commands. There the computation times were
comparable to those of determining a smallest critical command set
(cf.\ Table~\ref{tab:command}). Thus we suppose that the much smaller
times for simplification after selecting a smallest critical command
set are due to the considerably reduced possibilities for
simplification.

\begin{table}[tb]
  \caption{Results for branch and variable domain minimization (TO = 600 seconds)}
  \label{tab:upvar}
  \centering
  \scalebox{0.85}{\addtolength{\tabcolsep}{0.1pt}\begin{tabular}{lrrrrcrrrr}
  \toprule
        & \multicolumn{4}{c}{branch minimization} & &\multicolumn{4}{c}{var.\ domain minimization} \\
  \cmidrule(lr){2-5}\cmidrule(lr){7-10}
  Model & Time & Mem. & $n$ & lb & & Time & Mem. & $n$ & lb \\
  \midrule
  coin2-1 & 3.78 & 16 & 11 & \opt && 0.39 & 10 & 15 & \opt \\
  coin2-2 & \TO & 3293 & 11 & \opt && \TO & 1142 & 19 & 17 \\
  coin2-4 & \TO & 1548 & 11 & 9 && \TO & 938 & 25 & 15 \\
  coin2-5 & \TO & 814 & 11 & 9 && \TO & 809 & 29 & 15 \\
  coin2-6 & \TO & 1186 & 11 & 9 && \TO & 515 & 32 & 13 \\
  csma2-2 & 0.46 & 15 & 39 & \opt && 0.33 & 18 & 92 & \opt \\
  csma2-4 & \TO & 135 & 70 & 69 && 3.16 & 102 & 120 & \opt \\
  fw01 & 0.11 & 11 & 27 & \opt && 0.14 & 14 & 342 & \opt \\
  fw04 & 0.31 & 26 & 30 & \opt && 0.33 & 32 & 342 & \opt \\
  fw10 & 0.81 & 58 & 29 & \opt && 1.16 & 72 & 342 & \opt \\
  fw15 & 1.37 & 88 & 27 & \opt && 2.30 & 112 & 342 & \opt \\
  wlan0-2 & 1.58 & 23 & 31 & \opt && 0.96 & 241 & 50  & \opt \\
  wlan2-1 & 0.11 & 30 & 8 & \opt && 1.01 & 260 & 25 & \opt \\
  wlan2-2 & \TO & 1533 & 49 & 43 && 5.52 & 251 & 67 & \opt \\
  wlan2-3 & \TO & 332 & 37 & 32 && \TO   & 865 & 51 & 40 \\
  wlan3-1 & 0.47 & 91 & 8 & \opt && 3.57 & 871 & 25 & \opt \\
  wlan3-2 & \TO & 1624 & 49 & 42 && 8.68 & 889 & 66 & \opt \\
  \bottomrule
\end{tabular}
}
\end{table}

The experiments show that is feasible to use MILP formulations for
counterexample computation, although solving the MILPs is costly and
often optimality of the result cannot be proven by the solver within
the given time limit. Additionally we can see that in all cases we
are able to reduce the number of commands and to simplify the
commands, in some cases considerably, compared to the original
\tool{PRISM} model.


\section{Conclusion}
\label{sec:conclusion}

We have presented a new type of counterexamples for probabilistic
automata which are described using a guarded command language: We
computed a smallest subset of the commands which alone induces an
erroneous system. This requires the solution of a mixed integer
linear program whose size is linear in the size of the state space of
the PA. State-of-the-art MILP solvers apply sophisticated techniques
to find small command sets quickly, but they are often unable to prove 
the optimality of their solution.

For the MILP formulation of the smallest critical labeling problem
we both need decision variables for the labels and for the
scheduler inducing the maximal reachability probabilities of the
subsystem. On the other hand, model checking can be executed
without any decision variables using a linear programming formulation.
We therefore coupled a MAXSAT solver with a model checker for 
PAs. For many benchmark instances this reduced the computation 
time significantly. First results on this alternative method have been
published in~\cite{dehnert-et-al-atva-2014}.

\bibliographystyle{splncs}
\bibliography{literature}

\clearpage

\appendix
\section*{Appendix}

\section{Proof of Theorem~\ref{th:np_complete}}
\label{app:proof_np}

Let $\PA=(S,\sinit,\Act,P)$ be a PA, $\LT:S\times\Act\times\Distr(S)\times S\partialto2^{\Lab}$ a
labeling function and $w:\Lab\to\mathbb{R}^{\geq 0}$ a weight function. Assume
the reachability property $\mathcal{P}_{\leq\lambda}(\finally{T})$ is violated.

\begin{thm}
  To decide whether there is a critical label set $\Lab'\subseteq\Lab$
  with $w(\Lab') \leq k$ is NP-complete.
\end{thm}

\proof
  That the decision problem is in NP is obvious: we can guess a label
  set $\Lab'\subseteq\Lab$ and verify in polynomial time by computing the
  reachability probability $\probmax(\sinit,\finally{T})$, that $w(\Lab')\leq k$ and that
  $\probmax(\sinit,\finally{T})>\lambda$.

  To prove the NP-hardness, we give a reduction from exact 3-cover (X3C)~\cite{GJ79},
  similar to \cite{CV10}. X3C is defined as follows:
  \begin{quote}
    Let $X$ be a finite set with $|X|=3r$ and a collection $C\subseteq 2^X$ of
    subsets with $|c|=3$ for all $c\in C$. Is there a collection 
    of pairwise disjoint sets $B\subseteq C$ such that $X=\bigcup_{c\in B}c$?
  \end{quote}
  We note that $X$ has an exact 3-cover iff it has a cover of size $|B|\leq r$.

  Given $X$ and $C$, we construct a PA $\PA=(S,\sinit,\{\act\}, P)$ as follows: $S=X\dcup C\dcup \{\sinit,t\}$
  with two fresh states $\sinit$ and $t$. We set
  $P(\sinit) = \bigl\{ (\act,\mu_X)\bigr\}$ with $\mu_X(x)=\frac{1}{|X|}$ for all $x\in X$ and
  $\mu_X(s)=0$ for all $s\in S\setminus X$.
  For $x\in X$ we define $P(x) = \bigl\{ (\act,\mu^1_c)\,\big|\, c\in C\text{ with }x\in c\bigr\}$, where
  $\mu_c^1(c)=1$ and $\mu_c^1(s)=0$ for all $s\neq c$.
  Finally, $P(c)=\bigl\{ (\act,\mu^1_t) \bigr\}$ for all $c\in C$, and $P(t)=\emptyset$.

  We use the following label set $\Lab = \{\ell_{\sinit}\}\cup \{ \ell_x\,|\,x\in X\}
  \cup \{l_t\}$ and label the branches as follows: all branches in $P(\sinit)$ are
  labeled with $\ell_{\sinit}$, the branches in $P(x)$ with $\ell_x$ for $x\in X$
  and the branches in $P(c)$ with $\ell_t$ for $c\in C$.
  The weight function $w$ assigns $0$ to $\ell_{\sinit}$ and $\ell_t$, and 
  $1$ to all other labels. The set of target states is $\{t\}$, and the 
  probability bound $\lambda = 1-\frac{1}{|X|+1}$.

  We claim that there is a critical label set with weight $\leq r$ iff $X$ has an
  exact 3-cover.
  \begin{itemize}[label=``$\Rightarrow$'']
  \item[``$\Rightarrow$''] Let $\Lab'\subseteq\Lab$ be a critical label set with
    $w(\Lab')\leq r$. We can assume w.\,l.\,o.\,g.\ that $\ell_{\sinit}$ and
    $\ell_t$ are contained in $\Lab'$ since their weight is zero. We observe that
    from each state $x\in X$ there has to be a path to state $t$ in the induced
    sub-PA $\PA_{|\Lab'}$. Otherwise the maximal probability to reach $t$ from $\sinit$ is
    $\leq 1-\frac{1}{|X|} < 1-\frac{1}{|X|+1}$. This means, for each $x\in X$ there is $c\in C$ with
    $x\in c$ and $\ell_c\in\Lab'$. Let $B = \{c\in C\,|\,\ell_c\in\Lab'\}$.
    $B$ is a cover of $X$. Since $w(\Lab')\leq r$, $|B|\leq r$ and $B$ is an exact cover.
  \item[``$\Leftarrow$''] Let $B\subseteq C$ be an exact cover of $X$. We set
    $\Lab' = \{\ell_c\,|\,c\in B\}\cup\{\ell_{\sinit},\ell_t\}$. Then $w(\Lab')=r$.
    For all $x\in X$ there is $c\in B$ such that $x\in c$, because $B$ is a cover.
    That means, for all $x\in X$ there is a path from $x$ to $t$ with probability $1$
    in $\PA_{|\Lab'}$. Since $\ell_{\sinit}\in\Lab'$, we have that $\probmax(\sinit,\finally{\{t\}}) = 1$.
    Hence, $\Lab'$ is critical.\qed
  \end{itemize}

\end{document}